\documentclass[12pt]{article}
\pdfoutput=1

\usepackage{amsmath}
\usepackage{color,booktabs}
\usepackage{graphicx}
\usepackage{array} 
\usepackage{subfig}
\usepackage{pdflscape}
\usepackage{afterpage}
\usepackage{authblk,etoolbox}
\usepackage{geometry}

\makeatletter
\patchcmd{\@maketitle}{center}{flushleft}{}{}
\patchcmd{\@maketitle}{center}{flushleft}{}{}

\def\maketitle{{%
  \renewenvironment{tabular}[2][]
    {\begin{flushleft}}
    {\end{flushleft}}
  \AB@maketitle}}
\makeatother

\begin{document}

\title{Estimating the association between blood pressure variability and cardiovascular disease: An application using the ARIC Study}

\author[1,2]{Jessica K.~Barrett\thanks{Corresponding author, email: jkb23@medschl.cam.ac.uk}}

\author[2,3]{Raphael Huille}

\author[4]{Richard Parker}

\author[5]{Yuichiro Yano}

\author[6]{Michael Griswold}

\affil[1]{MRC Biostatistics Unit, University of Cambridge, Cambridge, UK}
\affil[2]{Department of Public Health and Primary Care, University of Cambridge, Cambridge, UK}
\affil[3]{\'Ecole Nationale de la Statistique et de l'Administration \'Economique, Malakoff, France}
\affil[4]{School of Social and Community Medicine, University of Bristol, Bristol,UK}
\affil[5]{Department of Preventive Medicine, University of Mississippi Medical Center, Mississippi, USA}
\affil[6]{Center of Biostatistics and Bioinformatics, University of Mississippi Medical Center, Mississippi, USA}

\date{}

\maketitle
\vspace{-0.5cm}

\begin{abstract}
The association between visit-to-visit systolic blood pressure variability and cardiovascular events has recently received a lot of attention in the cardiovascular literature. But blood pressure variability is usually estimated on a person-by-person basis, and is therefore subject to considerable measurement error. We demonstrate that hazard ratios estimated using this approach are subject to bias due to regression dilution and we propose alternative methods to reduce this bias: a two-stage method and a joint model. For the two-stage method, in stage one repeated measurements are modelled using a mixed effects model with a random component on the residual standard deviation. The mixed effects model is used to estimate the blood pressure standard deviation for each individual, which in stage two is used as a covariate in a time-to-event model. For the joint model, the mixed effects sub-model and time-to-event sub-model are fitted simultaneously using shared random effects. We illustrate the methods using data from the Atherosclerosis Risk in Communities (ARIC) study.
\end{abstract}
\vspace{2cm}


\newpage

\section{Introduction}

Systolic blood pressure (SBP) is universally recognised as an important risk factor for cardiovascular disease (CVD) and is routinely included in risk scores for CVD risk prediction \cite{Dagostino2008,Hippisley-Cox2017DevelopmentStudy}. The prognostic value of SBP is primarily based on the mean of measurements over multiple outpatient visits, whereas substantially less attention has been given to the variability of SBP across visits (i.e., visit-to-visit SBP variability). However, there is increasing evidence of associations between greater visit-to-visit SBP variability and CVD outcomes in community-based and clinical settings \cite{Muntner2015,Parati2013,rothwell2010,Shimbo2012}. Therefore, in addition to assessing mean clinic SBP levels over time, measurements of visit-to-visit SBP variability may improve the accuracy of CVD risk prediction, which is crucial for the optimization of patient care. 

The standard deviation (SD), coefficient of variation (CV), average real variability (ARV), and SBP variability independent of the mean (VIM) across multiple visits have been widely used to quantify visit-to-visit SBP variability \cite{Muntner2015,Parati2013,rothwell2010,Shimbo2012,YanoVisit-to-VisitChallenge}. However, these measures are estimated on a person-by-person basis, and are therefore subject to considerable measurement error. This measurement error causes regression dilution bias in the estimated association between visit-to-visit SBP variability and CVD \cite{Hughes1993}. 

Mixed effects models have been proposed which allow the within-individual variability to differ between individuals. For example, Hedeker et al. introduced the mixed location scale model, where the within-individual variances may be assumed to depend on an additional random effect, as well as time-constant and time-varying covariates \cite{Hedeker2008AnData,Hedeker2012}. More recently Goldstein et al. have explored the mixed location scale model using a Bayesian framework in the context of growth curve models\cite{GoldsteinMultilevelFunction}.

We consider extensions of the mixed location scale model to estimate the association of SBP variability with CVD. Repeated SBP measurements are modelled using a mixed effects model with a random component on the within-individual standard deviation, thus allowing the borrowing of information across individuals. The model allows for each individual to have different within-individual variation, which describes the variability in their SBP measurements. We propose two methods for estimating the association of this SBP variability with CVD. The first is a two-stage method, where in stage one we obtain estimates of the usual level of SBP and the SBP variability for each individual from the mixed effects model, and in stage two these are included as covariates in a survival model for the time to the first CVD event. The second method is a joint model where the repeated SBP measurements and the time to first CVD event are modelled simultaneously using shared random effects \cite{Rizopoulos2012JointR}.     

There are advantages and disadvantages in the use of two-stage versus joint models. The heavy computational burden associated with simultaneous analysis of repeated measurements and time-to-event data makes the two-stage method a more feasible option for large data sets. The two-stage method could also be readily incorporated into a landmarking approach for dynamic risk prediction \cite{Houwelingen2012DynamicAnalysis}, for example in the dynamic model for CVD risk prediction recently proposed by Paige et al \cite{Paige2018}. However, two-stage methods for the analysis of repeated measurements and time-to-event data have been shown to lead to bias in estimated covariate effects compared to joint models \cite{Sweeting2011}. This bias is due to two factors: (1) regression dilution bias caused by residual measurement error in the estimates of random effects from the mixed effects model and (2) bias caused by informative truncation of the repeated measurements by the event of interest. In the two-stage approach this second source of bias can be removed by separating longitudinal follow-up from survival follow-up if data allows, as we describe in section \ref{sec:methods}.  

The outline of the paper is as follows. In section \ref{sec:methods} we describe the methods, including the naive method that has predominantly been used in the CVD literature, and our proposed two-stage and joint model methods. In section \ref{sec:simstudies} we compare the methods using simulation studies and explore the accuracy of all methods in a variety of scenarios. We apply the two-stage and joint models to data from the Atherosclerosis Risk in Communities (ARIC) Study in section \ref{sec:ARIC}, and we present a final discussion in section \ref{sec:discussion}.  

\section{Methods}
\label{sec:methods}

The association of SBP with CVD has previously been investigated using mixed effects models to analyse repeated measurements of SBP \cite{Sweeting2016TheStudy,Barrett2017DynamicStudy,Yang2016PredictionBiomarkers}. In this section we first present the naive methods for estimating SBP variability that have been commonly used in the literature, and then explore how the mixed effects model can be extended to allow SBP variability to differ between individuals. We propose a two-stage method and a joint modelling method for estimating the association between SBP variability and CVD events. Although we focus here on the association between SBP variability and CVD, the methods and results presented would be relevant for any application where the association between the variability in a longitudinal outcome and a time-to-event is of interest. We consider estimating the association of the variability of a longitudinal outcome with a time-to-event, adjusted for the `usual level' of the longitudinal outcome. This `usual level' could be the average value, which may be assumed constant in time, or the current value or baseline value if time-dependence is to be taken into account. 

\subsection{Notation}

Given a data set with $N$ individuals and $n_i$ SBP measurements from the $i$th individual, $i=1,\ldots,N$, let $Y_{ij}$, $j=1,\ldots,n_i$,  be the $j$th measurement of individual $i$ taken at measurement time $t_{ij}$. Each individual is followed up from baseline until time $T_i^* = \min (T_i, C_i)$, where $T_i$ is the time of the event for individual $i$ and $C_i$ is the censoring time.

\subsection{Estimating longitudinal variability}
\label{sec:BPV}

\subsubsection{Naive method}
\label{sec:naive}

In the cardiovascular literature the usual level of SBP is typically estimated as the mean of an individual's repeated measurements,
\[
\hat{BP}_{i}^{naive} = \frac{\sum_j Y_{ij}}{n_i}
\]
and the SBP variability as their standard deviation
\[
\widehat{BPSD}_i^{naive} = \sqrt{\frac{\sum_j (\hat{BP}^{naive}_i-Y_{ij})^2}{n_i}}
\] 
The repeated measurements may have been taken prior to follow-up of the time-to-event, or concurrent with the time-to-event follow-up, in which case the repeated measurement process is terminated by an event. Other measures of visit-to-visit SBP variability have been proposed, such as the coefficient of variation (defined as the SD/mean), the maximum and minimum difference and the average real variability (defined as the average change between successive visits) \cite{YanoVisit-to-VisitChallenge}, but all are usually estimated using within-individual information only. 

\subsubsection{Mixed effects models}
\label{sec:LMM}

An alternative method to estimate the usual level and variability for each individual is to use a mixed effects model for the repeated measurements from all individuals, allowing information to be borrowed across individuals. The mixed effects model also allows longitudinal trajectories to be modelled by including time-dependent terms in the model.

Consider a standard linear mixed effects model,
\begin{equation}
Y_{ij} = \beta^T X_{ij}  + b_i^T Z_{ij} + \epsilon_{ij} \,\, ,
\label{eq:Y}
\end{equation}
where $X_{ij}$ is a covariate vector for the fixed effects $\beta$ and $Z_{ij}$ is a covariate vector for the random effects $b_i$, assumed normally distributed $b_i \sim N(0, \Sigma_b)$. The residual errors $\epsilon_{ij}$ are assumed independent and normally distributed, $\epsilon_{ij} \sim N(0, \sigma^2)$. We can allow variability in the repeated measurements to differ between individuals by replacing the residual standard deviation (SD) $\sigma$ with an individual-specific residual SD $\sigma_i$ and assuming that the $\sigma_i$ are randomly distributed. Note that in this model we do not distinguish between true variability in an individual's repeated measurements and measurement error. We assume a log-Normal distribution for the residual SD distribution, ensuring positivity of the SDs, $\sigma_i \sim \text{logN} (\mu_\sigma , \tau_\sigma^2)$. The choice of log-Normal distribution also allows a natural extension of the model to incorporate correlation between the usual level and the residual SD by assuming a multivariate normal distribution for the random effects and log residual SD,
\begin{equation}
\epsilon_{ij} \sim N(0,\sigma_i^2) \,\, , \quad
\begin{pmatrix} b_i \\ \log\sigma_i \end{pmatrix} \sim N \left( 
\begin{pmatrix} 0 \\ \mu_{\sigma} \end{pmatrix} ,
\begin{pmatrix} \Sigma_b & \Sigma_{b \sigma} \\ \Sigma^T_{b \sigma} & \tau^2_{\sigma} \end{pmatrix}
\right) \,\, ,
\label{eq:sigma}
\end{equation}
where $\Sigma_{b \sigma}$ is a vector of covariances between the random effects and the random residual errors. Alternative distributions have been proposed for the residual SD, such as the half-Cauchy distribution \cite{LunnTheAnalysis}. (See also Hedeker et al \cite{Hedeker2008AnData} and recent work by Goldstein et al \cite{GoldsteinMultilevelFunction}, where a log-Normal distribution was assumed for the residual variances $\sigma_i^2$.)

Specific examples of the model specified by equations \eqref{eq:Y} and \eqref{eq:sigma} which we will consider are a random-intercept model
\begin{equation}
Y_{ij} = \beta^T X_{i} + b_{0i} + \epsilon_{ij} \,\, , 
\label{eq:Y_randint}
\end{equation}
where the random intercept $b_{0i}$ can be interpreted as the usual level of the repeated measurements, and adjustment is made for baseline covariates $X_i$. We consider a model (LMM1) which does not account for the correlation between $b_0i$ and $\sigma_i$,
\begin{equation}
\text{LMM1: } 
\begin{pmatrix} b_{0i} \\ \log\sigma_i \end{pmatrix} \sim N \left( 
\begin{pmatrix} 0 \\ \mu_{\sigma} \end{pmatrix} ,
\begin{pmatrix} \tau_0^2 & 0 \\ 0 & \tau^2_{\sigma} \end{pmatrix} \,\, , 
\right)
\label{eq:LMM1}
\end{equation}
and a model (LMM2) which does account for the correlation
\begin{equation}
\text{LMM2: } 
\begin{pmatrix} b_{0i} \\ \log\sigma_i \end{pmatrix} \sim N \left( 
\begin{pmatrix} 0 \\ \mu_{\sigma} \end{pmatrix} ,
\begin{pmatrix} \tau_0^2 & \rho \tau_0 \tau_\sigma \\ \rho \tau_0 \tau_\sigma & \tau^2_{\sigma} \end{pmatrix} \,\, .
\right)
\label{eq:LMM2}
\end{equation}
We also consider a random intercept and slope model
\[
Y_{ij} = X_{i} \beta + \beta_t t_{ij} + b_{0i} + b_{1i} t_{ij} + \epsilon_{ij} \,\, , 
\]
where the random intercept $b_{0i}$ is now the baseline value (minus the population average), and $\sigma_i$ now measures the variability around the individual's average trajectory. Again, we can either ignore or allow for correlations between the random effects and the residual SD's,
\begin{equation}
\begin{split}
\text{LMM3: } &
\begin{pmatrix} b_{0i} \\ b_{1i} \\ \log\sigma_i \end{pmatrix} \sim N \left( 
\begin{pmatrix} 0 \\ 0 \\ \mu_{\sigma} \end{pmatrix} ,
\begin{pmatrix} \tau_0^2 & \rho_{01} \tau_0 \tau_1 & 0 \\ 
\rho_{01} \tau_0 \tau_1 & \tau_1^2 & 0 \\
0 &   0 & \tau^2_{\sigma} \end{pmatrix} 
\right)
\,\, , \\[0.3cm] 
\text{LMM4: } &
\begin{pmatrix} b_{0i} \\ b_{1i} \\ \log\sigma_i \end{pmatrix} \sim N \left( 
\begin{pmatrix} 0 \\ 0 \\ \mu_{\sigma} \end{pmatrix} ,
\begin{pmatrix} \tau_0^2 & \rho_{01} \tau_0 \tau_1 & \rho_{0\sigma} \tau_0 \tau_\sigma \\ 
\rho_{01} \tau_0 \tau_1 & \tau_1^2 & \rho_{1\sigma} \tau_1 \tau_\sigma \\
\rho_{0\sigma} \tau_0 \tau_\sigma &   \rho_{1\sigma} \tau_1 \tau_\sigma & \tau^2_{\sigma} \end{pmatrix} 
\right) \,\, .
\end{split}
\end{equation}

\subsection{Estimating the association between SBP variability and CVD}

We consider two approaches for estimating the association between SBP variability and the time to the first CVD event, a two-stage approach and a joint model. 

\subsubsection{Two-stage approach}
\label{sec:two-stage}
We fit the models in two stages. In stage one we either (i) estimate the usual level and SD of the repeated measurements for each individual using the naive method, or (ii) fit a mixed effects model and estimate the usual level, residual SD and other random effects for each individual, as described in section \ref{sec:BPV}. In the second stage we use the estimated usual level and residual SD as covariates in a standard proportional hazards Cox regression for the time to the first event,
\[
h_i(t) = h_0(t) \exp (\alpha^T U_i + \gamma^T W_i) \,\, ,
\]
where $U_i$ is a vector containing the estimated usual level, residual SD and other random effects for each individual, $\alpha$ is a vector of parameters describing the association between the mixed effects model and the time-to-event model and $W_i$ is a vector of baseline covariates.  

We estimate mixed effects model parameters using Bayesian MCMC implemented using JAGS Version 3.4.0 \cite{jags2017} and the {\tt R2jags} package in {\tt R} \cite{R2jags}. We assume diffuse priors for all model parameters. The usual level $\hat{BP}_i^{LMM}$ is estimated by the posterior mean of the random intercept $b_{0i}$ and the variability $\widehat{BPSD}_i^{LMM}$ by the posterior mean of the residual SD $\sigma_i$. Additional random effects can also be estimated by posterior means.   

The SBP measurement process is truncated informatively by the event of interest, as illustrated in Figure ~\ref{fig:1a}. In order to avoid this informative truncation we can divide all follow-up into follow-up of the repeated measurements and follow-up of the survival outcome, separated at separation time $t_{sep}$, as shown in Figure \ref{fig:1b}. Only repeated measurements taken before $t_{sep}$ are used in estimation of the mixed effects model, and only events taking place after the separation time are used in estimation of the time-to-event model conditional on surviving until $t_{sep}$. Separating the longitudinal follow-up from the survival follow-up in this way involves discarding information from repeated measurements following $t_{sep}$ and individuals experiencing events before $t_{sep}$. In practice $t_{sep}$ should be chosen to minimise this information loss.

\subsubsection{Joint model}
\label{sec:joint}
In the joint model we use a shared parameter approach to link the mixed effects and time-to-event models. The mixed effect sub-model is defined by equations \eqref{eq:Y} and \eqref{eq:sigma}. The time-to-event sub-model is defined by
\[
h_i(t) = h_0(t) \exp (\alpha^T b_i + \gamma^T W_i) \,\, .
\]
where $b_i$ are the random effects shared with the mixed effects sub-model. We assume a piecewise constant baseline hazard; we choose cut-points $t_k$, $0 \leq k \leq K$, at the K-quantiles of the observed event times and assume that the baseline hazard is constant in the time intervals between cut-points, $h_0(t) = h_{0,k}$, $t_{k-1}< t \leq t_k$. The piecewise-constant assumption ensures that the hazard function can be integrated analytically, thus avoiding the need for numerical integration when evaluating the likelihood function. In practice the number of cut-points can be chosen by assessing model fit, for example using the deviance information criterion (DIC)\cite{DIC}.

We again estimate model parameters using Bayesian MCMC implemented using JAGS Version 3.4.0 \cite{jags2017} and the {\tt R2jags} package in {\tt R} \cite{R2jags} and assuming diffuse priors for all model parameters. 

\subsection{Regression dilution bias in association parameters}

In the two-stage approach, because SBP variability is estimated with error we expect bias in the estimated association with CVD due to regression dilution. We aim to reduce regression dilution bias by using a mixed effects model which allows information to be borrowed across individuals. The joint model accounts for this measurement error by including the underlying random effects in the time-to-event model, rather than estimates of those random effects. However, misspecification of the mixed or joint models could introduce other sources of bias. In Section 1 of the Supplementary Materials we show for the simplified case of linear regression that ignoring the correlation between the random effects and the residual SD could also lead to bias in the estimated effects of both the usual level and the variability on the outcome.

\begin{figure}[h]
	\centering
	\subfloat[]{\includegraphics[width=0.47\textwidth]{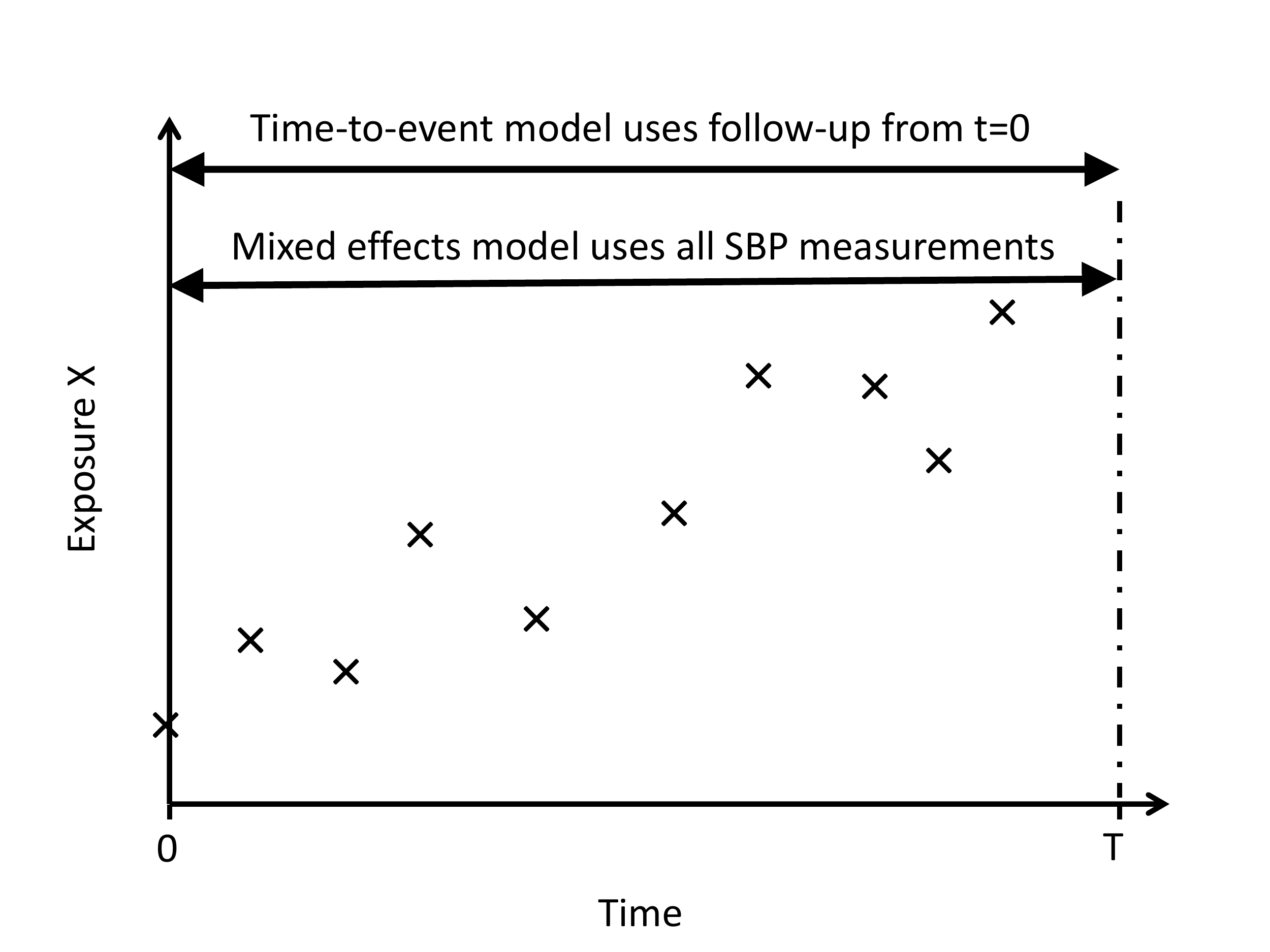}\label{fig:1a}} \qquad 
	\subfloat[]{\includegraphics[width=0.47\textwidth]{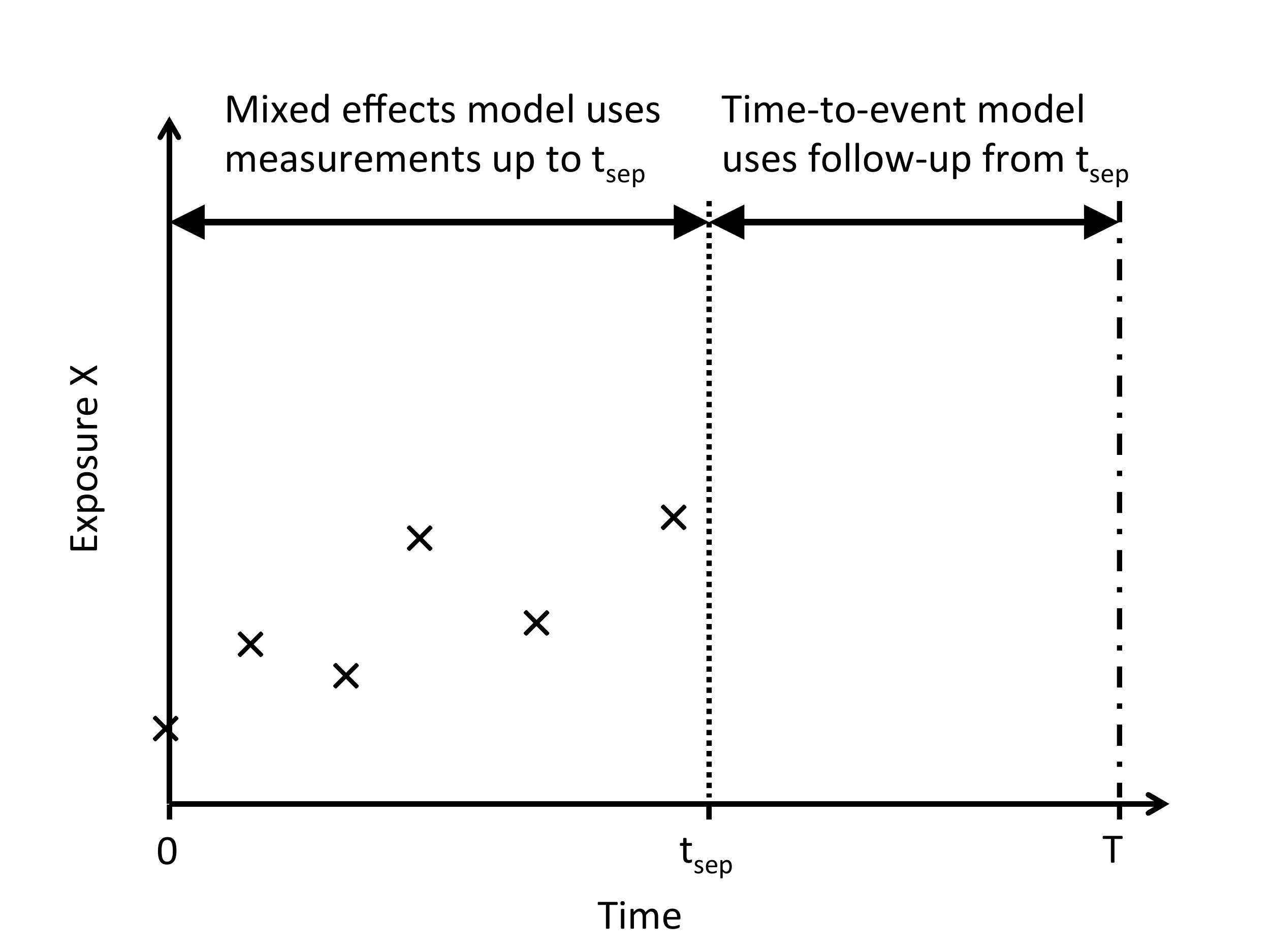}\label{fig:1b}} \qquad 
	\caption{Diagrams illustrating the use of SBP measurements and time-to-event data when (a) time-to-event follow-up and SBP measurement follow-up run concurrently from time t=0 and (b) repeated measurement follow-up takes place prior to time-to-event follow-up.}
    \label{fig:sepdiag}
\end{figure}


\section{Simulation studies}
\label{sec:simstudies}

We use simulation studies to explore the bias in parameter estimates due to imprecise estimation of SBP variability, and due to ignoring correlation between SBP and SBP variability in the mixed effects model. We consider two scenarios. In Scenario 1 each individual has the same number of repeated measurements, thus bias in the two-stage approaches arises solely from measurement error in the estimates of the usual level and variability of SBP (c.f. Figure \ref{fig:1b}). In Scenario 2 the repeated measurement process is truncated by the event times, so that bias in the two-stage approaches arises from both measurement error and informative truncation of the repeated measurements (c.f. Figure \ref{fig:1a}). We generated simulated data sets to explore the effect of the number of measurements per individual, the value of the association parameters and the extent of the correlation between the usual level and the variability.

We take a simple model for the repeated measurements with a random intercept only and random residual error standard deviation
\[
BP_{ij} = b_{0i} + \epsilon_{ij}
\]
where the residual error $\epsilon_{ij} \sim N(0, \sigma_i^2)$ and we take a bivariate normal distribution for $b_{0i}$ and $ \log\sigma_i$,
\[
\begin{pmatrix} b_{0i} \\ \log\sigma_i \end{pmatrix} \sim N \left( 
\begin{pmatrix} \mu_0 \\ \mu_{\sigma} \end{pmatrix} ,
\begin{pmatrix} \tau_0^2 & \rho \tau_0\tau_\sigma \\  \rho \tau_0\tau_\sigma & \tau^2_{\sigma} \end{pmatrix}
\right) \,\, ,
\]
Here $b_{0i}$ is the usual level and $\sigma_i$ represents the variability in the longitudinal outcome. Based on results from the ARIC study, we take $\mu_0=120$, $\tau_0=15$, $\mu_{\sigma}=2$ and $\tau_\sigma=0.5$. Event times are drawn from a Weibull distribution with shape parameter $k$ and scale parameter $\lambda$. A loglinear effect of the covariates on $\lambda$ gives a proportional hazards model,
\[
\lambda = \exp (\gamma_0 + \alpha_0 b_{0i} + \alpha_\sigma \sigma_i)
\]
We introduce administrative censoring at 20 years. To maintain an event rate of 20\% before censoring we take $\gamma_0=-10.26$ and $k=2$ with default association parameters $\alpha_0=0.02$ and $\alpha_\sigma=0.05$, similar to those found for the ARIC data set. For Scenario 1 we used all repeated measurements for all individuals. For Scenario 2 we assumed repeated measurements were taken equidistantly between baseline and 18 years follow-up and discarded all repeated measurements following event times. In each simulation set-up we generated 1000 data sets, each consisting of 1500 individuals. 

We analysed each generated data set using the two-stage method from section \ref{sec:two-stage}, with (i) naive and (ii) mixed model estimates of SBP usual levels and variabilities, and the joint models from section \ref{sec:joint}. For the mixed effects and joint models we fitted models ignoring the correlation between SBP and SBP variability (mixed effects model LMM1 and joint model JM1 used equations \eqref{eq:Y_randint} and \eqref{eq:LMM1}), and allowing for the correlation (mixed effects model LMM2 and joint model JM2 used equations \eqref{eq:Y_randint} and \eqref{eq:LMM2}). The joint models were fitted using 15 time intervals for the baseline hazard. Model convergence was checked using the Gelman-Rubin statistic as modified by Brooks and Gelman \cite{Brooks1998GeneralSimulations}. We also fitted the survival models using the true values of $b_{0i}$ and $\sigma_i$ as covariates, representing the best achievable results from the time-to-event model when all covariates are known precisely. For all scenarios we calculated the mean and standard deviation of the estimated log hazard ratios and the root mean squared errors (RMSE) and coverage probabilities at the 95\% level.

Table \ref{table:n} shows results for $n=4$, $n=7$ and $n=10$ measurements per individual for Scenario 1. With 4 measurements per individual using the naive method leads to a slight bias in the log hazard ratio (logHR) for the usual level, but substantial negative bias and associated loss of coverage in the logHR for variability, which is most likely caused by regression dilution due to imprecise measurement of longitudinal variability. Using the  mixed effects model LMM1 leads to a positive bias in both the usual level logHR and the variability logHR, as we would expect from the arguments given in Section1 of the Supplementary Materials, because correlation between these quantities was not accounted for in their estimation. Using method LMM2, which accounts for the correlation, results in minimal bias, even with only 4 measurements per individual. For both two-stage models the coverage is a little below the nominal level, but the RMSE is larger for the variability logHR in LMM2 due to greater variation in effect estimates. Results for the joint models are similar to the two-stage models in terms of the bias, but coverage is closer to the nominal levels. Results for n=7 and n=10 are similar, with the naive model giving reduced but still substantial bias in the variability logHR with a greater number of measurements. 

Table \ref{table:ntrunc} shows results for $n=4$, $n=7$ and $n=10$ measurements per individual for Scenario 2. The two-stage model LMM2 now also gives negative bias in estimating the variability logHR due to informative truncation of the repeated measurements by the CVD event. The joint model JM2, which accounts for this informative truncation, gives consistent results. Our results suggest that the bias incurred in the two-stage approach by informative truncation is more considerable here than the bias incurred by error in the estimates of SBP variability from the LMM models.  

\afterpage{
\begin{landscape}
\begin{table}[h]
\caption{Scenario 1 results with different numbers of repeated measurements. Presented are the true values, mean (standard deviation) of estimated log hazard ratios, root mean squared error and coverage over 1000 simulated datasets. Methods of analysis are (1) True values, where the true usual levels and variabilities are used as covariates, (2) Naive method, (3) LMM1, the two-stage approach with no correlation between the usual level and the variability, (4) LMM2, the two-stage approach with correlation, (5) JM1, the joint model with no correlation and (6) JM2, the joint model with correlation.  \label{table:n}}
\centering
\begin{tabular}{lllllcllll}
\toprule
\multicolumn{1}{l}{\bfseries  }&\multicolumn{4}{c}{\bfseries Usual level logHR \boldmath$\alpha_0$}&\multicolumn{1}{c}{\bfseries }&\multicolumn{4}{c}{\bfseries Variability logHR \boldmath$\alpha_\sigma$} \\
\cline{2-5} \cline{7-10}
&\multicolumn{1}{c}{\bfseries True}&\multicolumn{1}{c}{\bfseries Mean (SD)}&\multicolumn{1}{c}{\bfseries RMSE}&\multicolumn{1}{c}{\bfseries Coverage}&\multicolumn{1}{c}{}&\multicolumn{1}{c}{\bfseries True}&\multicolumn{1}{c}{\bfseries Mean (SD)}&\multicolumn{1}{c}{\bfseries RMSE}&\multicolumn{1}{c}{\bfseries Coverage}\\ \midrule
{\bfseries n=4}&&&&&&&&&\\
~~True values&0.02&0.02 (0.0043)&0.0043&0.94&&0.05&0.0501 (0.0117)&0.0117&0.94\\
~~Naive&0.02&0.0208 (0.0037)&0.0038&0.947&&0.05&0.0284 (0.0094)&0.0235&0.31\\
~~LMM1&0.02&0.0227 (0.0042)&0.005&0.892&&0.05&0.0531 (0.0151)&0.0154&0.929\\
~~LMM2&0.02&0.0198 (0.0049)&0.0049&0.948&&0.05&0.0511 (0.0175)&0.0175&0.931\\
~~JM1&0.02&0.023 (0.0042)&0.0052&0.883&&0.05&0.0521 (0.0155)&0.0156&0.939\\
~~JM2&0.02&0.02 (0.0051)&0.0051&0.949&&0.05&0.0502 (0.0183)&0.0183&0.936\\
{\bfseries n=7}&&&&&&&&&\\
~~True values&0.02&0.0202 (0.0042)&0.0042&0.946&&0.05&0.0496 (0.0113)&0.0113&0.951\\
~~Naive&0.02&0.0209 (0.0039)&0.004&0.943&&0.05&0.0354 (0.01)&0.0177&0.684\\
~~LMM1&0.02&0.022 (0.0042)&0.0047&0.913&&0.05&0.0519 (0.0133)&0.0134&0.943\\
~~LMM2&0.02&0.0202 (0.0046)&0.0046&0.95&&0.05&0.0502 (0.0146)&0.0146&0.944\\
~~JM1&0.02&0.0221 (0.0042)&0.0048&0.91&&0.05&0.0509 (0.0135)&0.0136&0.952\\
~~JM2&0.02&0.0203 (0.0047)&0.0047&0.945&&0.05&0.0493 (0.015)&0.015&0.946\\
{\bfseries n=10}&&&&&&&&&\\
~~True values&0.02&0.0199 (0.0041)&0.0041&0.953&&0.05&0.05 (0.0111)&0.0111&0.956\\
~~Naive&0.02&0.0206 (0.004)&0.004&0.948&&0.05&0.039 (0.0102)&0.015&0.818\\
~~LMM1&0.02&0.0212 (0.0042)&0.0044&0.938&&0.05&0.0516 (0.0124)&0.0125&0.95\\
~~LMM2&0.02&0.0199 (0.0045)&0.0045&0.944&&0.05&0.0501 (0.0134)&0.0134&0.95\\
~~JM1&0.02&0.0213 (0.0042)&0.0044&0.931&&0.05&0.0508 (0.0125)&0.0125&0.956\\
~~JM2&0.02&0.02 (0.0045)&0.0045&0.943&&0.05&0.0494 (0.0136)&0.0136&0.946\\
\bottomrule  
\end{tabular}
\end{table}
\end{landscape}
}

\afterpage{
\begin{landscape}
\begin{table}[h]
\caption{Scenario 2 results with different numbers of repeated measurements. Presented are the true values, mean (standard deviation) of estimated log hazard ratios, root mean squared error and coverage over 1000 simulated datasets. Methods of analysis are (1) True values, where the true usual levels and variabilities are used as covariates, (2) Naive method, (3) LMM1, the two-stage approach with no correlation between the usual level and the variability, (4) LMM2, the two-stage approach with correlation, (5) JM1, the joint model with no correlation and (6) JM2, the joint model with correlation. \label{table:ntrunc}}
\centering
\begin{tabular}{lllllcllll}
\toprule
\multicolumn{1}{l}{\bfseries  }&\multicolumn{4}{c}{\bfseries Usual level logHR \boldmath$\alpha_0$}&\multicolumn{1}{c}{\bfseries }&\multicolumn{4}{c}{\bfseries Variability logHR \boldmath$\alpha_\sigma$} \\
\cline{2-5} \cline{7-10}
&\multicolumn{1}{c}{\bfseries True}&\multicolumn{1}{c}{\bfseries Mean (SD)}&\multicolumn{1}{c}{\bfseries RMSE}&\multicolumn{1}{c}{\bfseries Coverage}&\multicolumn{1}{c}{}&\multicolumn{1}{c}{\bfseries True}&\multicolumn{1}{c}{\bfseries Mean (SD)}&\multicolumn{1}{c}{\bfseries RMSE}&\multicolumn{1}{c}{\bfseries Coverage}\\ \midrule
{\bfseries n=4}&&&&&&&&&\\
~~True values&0.02&0.02 (0.0042)&0.0042&0.938&&0.05&0.0496 (0.0115)&0.0115&0.942\\
~~Naive&0.02&0.0229 (0.0041)&0.005&0.86&&0.05&0.0118 (0.0122)&0.0401&0.036\\
~~LMM1&0.02&0.0229 (0.0042)&0.0051&0.892&&0.05&0.0348 (0.0153)&0.0216&0.868\\
~~LMM2&0.02&0.0203 (0.0047)&0.0047&0.952&&0.05&0.0385 (0.0166)&0.0202&0.934\\
~~JM1&0.02&0.0236 (0.0047)&0.0059&0.852&&0.05&0.0409 (0.0211)&0.0229&0.908\\
~~JM2&0.02&0.0199 (0.0055)&0.0055&0.94&&0.05&0.0493 (0.022)&0.022&0.944\\
{\bfseries n=7}&&&&&&&&&\\
~~True values&0.02&0.0203 (0.004)&0.0041&0.96&&0.05&0.0498 (0.011)&0.011&0.964\\
~~Naive&0.02&0.0225 (0.0039)&0.0047&0.89&&0.05&0.0242 (0.011)&0.028&0.28\\
~~LMM1&0.02&0.0221 (0.004)&0.0045&0.92&&0.05&0.0418 (0.0126)&0.015&0.928\\
~~LMM2&0.02&0.0205 (0.0042)&0.0042&0.96&&0.05&0.0437 (0.0133)&0.0147&0.96\\
~~JM1&0.02&0.0223 (0.0043)&0.0049&0.904&&0.05&0.0466 (0.0153)&0.0157&0.952\\
~~JM2&0.02&0.0202 (0.0046)&0.0046&0.952&&0.05&0.0502 (0.0156)&0.0156&0.968\\
{\bfseries n=10}&&&&&&&&&\\
~~True values&0.02&0.0203 (0.0041)&0.0041&0.954&&0.05&0.0506 (0.012)&0.012&0.938\\
~~Naive&0.02&0.0222 (0.004)&0.0045&0.902&&0.05&0.031 (0.0121)&0.0226&0.544\\
~~LMM1&0.02&0.0216 (0.004)&0.0043&0.93&&0.05&0.0449 (0.0136)&0.0145&0.918\\
~~LMM2&0.02&0.0205 (0.0042)&0.0042&0.958&&0.05&0.0458 (0.0139)&0.0145&0.94\\
~~JM1&0.02&0.0216 (0.0043)&0.0045&0.926&&0.05&0.0487 (0.0155)&0.0155&0.918\\
~~JM2&0.02&0.0202 (0.0045)&0.0045&0.956&&0.05&0.0506 (0.0159)&0.0159&0.932\\
\bottomrule  
\end{tabular}
\end{table}
\end{landscape}
}

In Section 2 of the Supplementary Materials we present further results investigating performance of the models for different levels of association between the longitudinal trajectories and the CVD event (Supplementary Table 1) and for different levels of correlation between the usual level and the variability (Supplementary Table 2). All results are given for datasets with $n=4$ measurements. In brief, all methods performed well with no association between times-to-events and longitudinal variability. But when the associations between the time-to-event and the usual level and variability of the longitudinal trajectories was substantially stronger, all the methods struggled to give consistent results with only 4 repeated measurements. The joint models performed the best in this case with the least bias and highest coverage (Supplementary Table 1). When there is negative correlation between usual levels and variabilities the direction of bias is reversed. Bias in the usual level logHR using methods LMM1 and JM1 increases with increasing $\rho$. But for the variability logHR the pattern is less clear because of the interplay between regression dilution bias and the bias incurred by ignoring correlation in the mixed effects model (Supplementary Table 2). 

In summary, the method which overall resulted in the least bias was JM2. For parameter values similar to those observed in the data example method LMM2 performed well. But for all methods biases were observed with strong association parameters with $n=4$ measurements per individual, so more measurements per individual would be required in this scenario. The interplay between biases due to multiple causes is complex, and depends on the correlation parameter $\rho$. The RMSE, however, is generally slightly higher for LMM2 than for LMM1 and for JM2 than for JM1, suggesting a trade-off between the bias caused by ignoring the correlation for models LMM1 and JM1 and the higher variance induced by the increased complexity of models LMM2 and JM2.

\section{Example: ARIC study}
\label{sec:ARIC}

We illustrate our methods using data from the ARIC study \cite{aric1989}. Briefly, 15,792 mostly black and white adults aged 45-64 years were enrolled into the ARIC study between 1987 and 1989 via probability sampling from 4 U.S. communities: Washington County, Maryland; Forsyth County, North Carolina; Minneapolis, Minnesota, suburbs; and Jackson, Mississippi. Participants underwent five examinations during 25-years of follow-up (i.e., Visit 1, 2, 3, 4, and 5 examinations), with an annual contact by telephone. In the current analysis, we used SBP measurements from Visit 1 (1987-1989) through Visit 4 (1996-1998). 

We analysed both the full ARIC dataset and a reduced dataset where longitudinal follow-up was truncated at the last systolic SBP measurement taken at Visit 4 (at $t_{sep}=11.9$ years from baseline) and the time origin for the survival follow-up was taken at the same time-point (Figure \ref{fig:1b}). For the reduced dataset, to have enough measurements to estimate the standard deviation of repeated measurements using the naive method we restricted to individuals with three or four non-missing measurements recorded and also to individuals who did not experience an event before $t_{sep}$. In total, our full data set consisted of 13,161 individuals and our reduced data set consisted of 10,019 individuals. Supplementary Table 3 shows the baseline characteristics of the reduced and full datasets.

\afterpage{
\begin{figure}[h]
\centering
\subfloat[]{\includegraphics[width=0.47\textwidth]{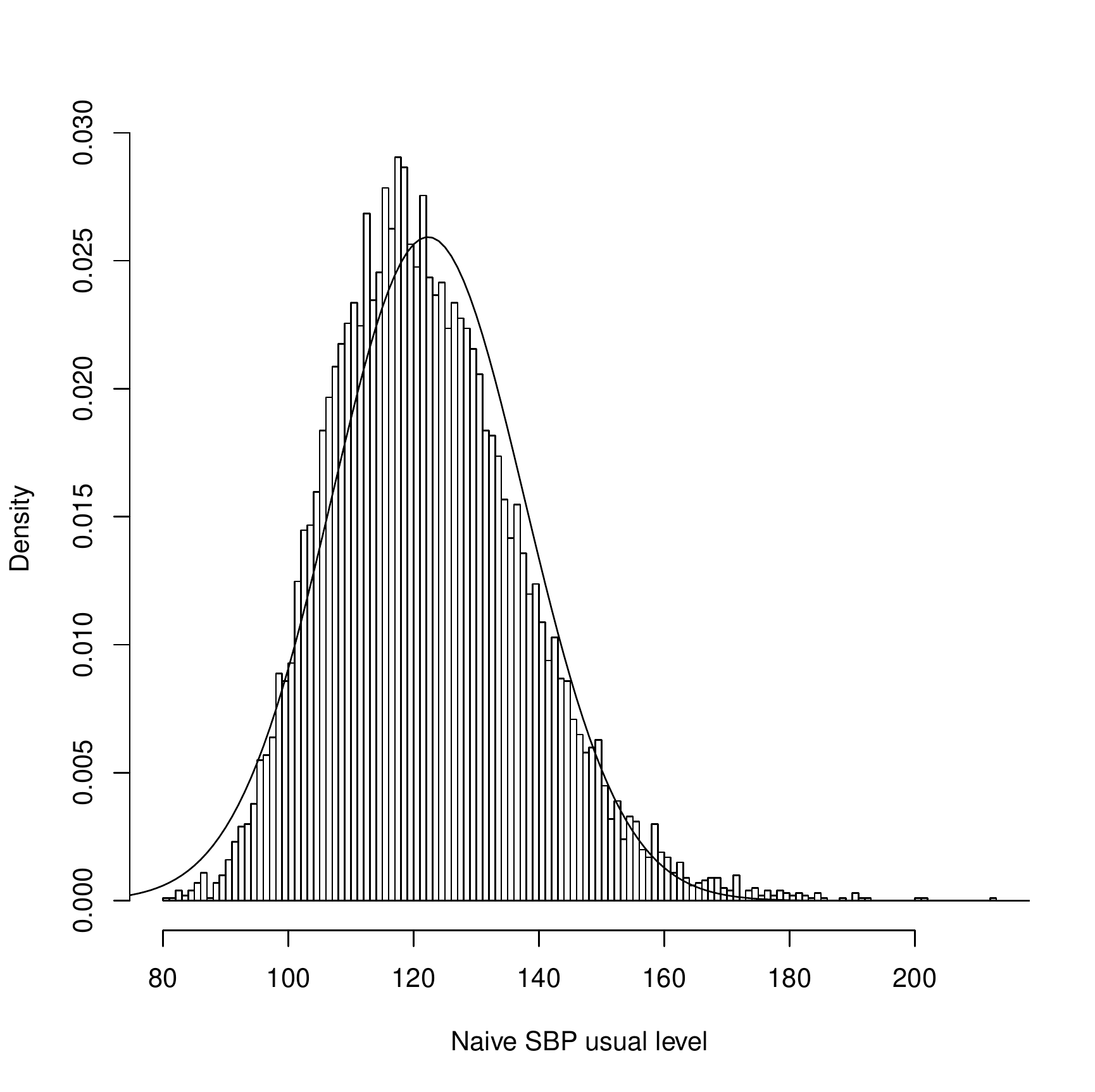}} \qquad
\subfloat[]{\includegraphics[width=0.47\textwidth]{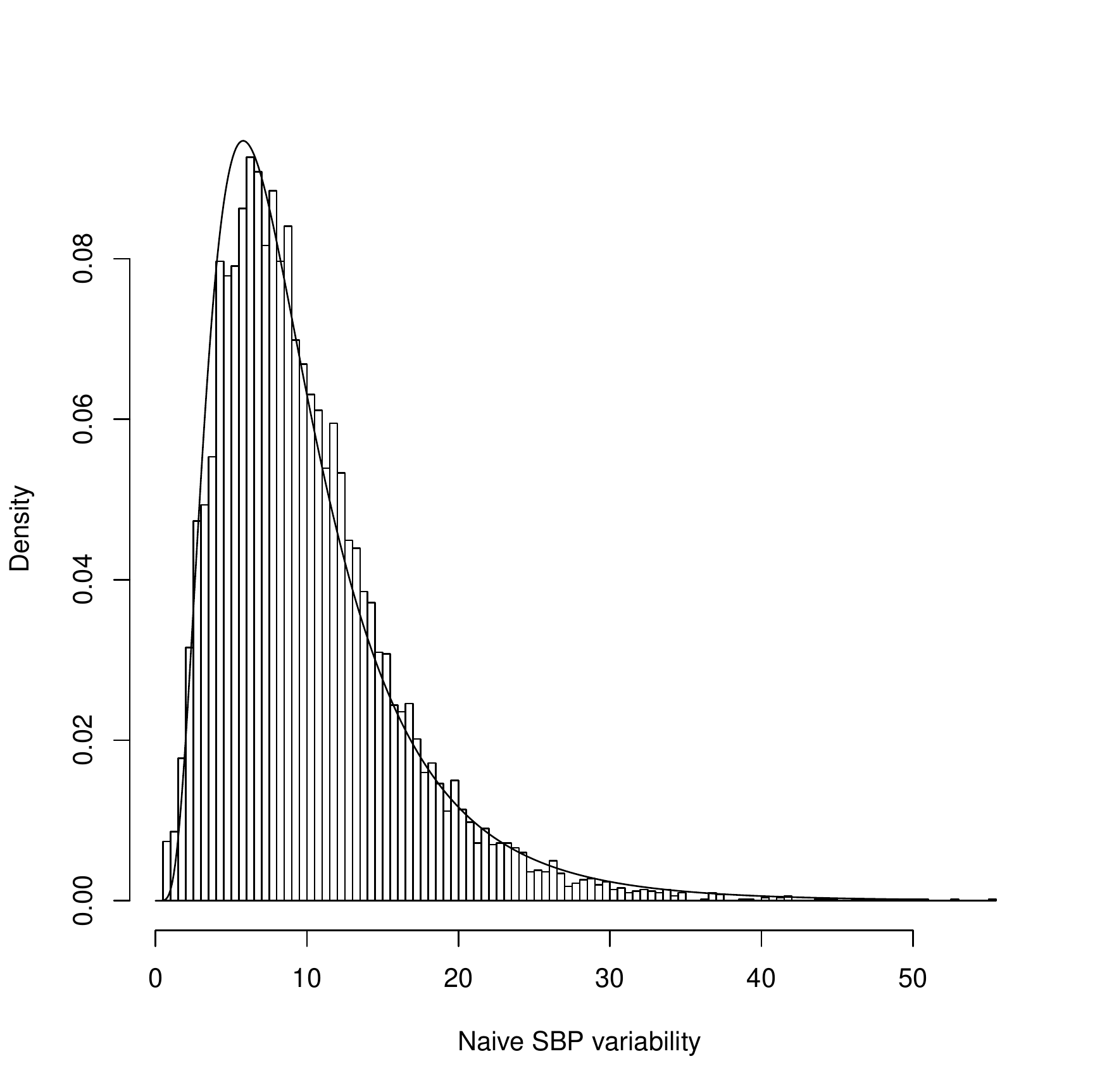}}
    \caption{Histograms of the distribution of (a) naive usual SBP estimates and (b) naive SBP variability estimates in data from the ARIC study. Also plotted are probability density functions of (a) a normal distribution fitted to the data and (b) a log-normal distribution fitted to the data.}
    \label{fig:hist}
\end{figure}
}

We analysed both the full and reduced data-sets using two-stage and joint models. For the two-stage approaches we used the repeated measurements of SBP to estimate the usual SBP level and the SBP variability using the naive method and the linear mixed effects models without and with accounting for the correlation between SBP and SBP variability (we again label these models LMM1 and LMM2 respectively). We also fitted joint models both without and with correlation between SBP and SBP variability (again JM1 and JM2 respectively). We applied the mixed effects models/sub-models in four ways
\begin{enumerate}
\item With a random intercept only, not adjusting for baseline CVD risk factors
\item Including a slope term in the model as both a fixed and random effect. The covariance matrix between the random effects and the residual log-standard deviation was taken to be unstructured, allowing for correlations between the random slopes and the random intercept and between the random slopes and the SBP variability.
\item Adjusting the random intercept mixed effects model for baseline CVD risk factors (age, sex, diabetes status, smoking status, baseline total cholesterol and baseline HDL cholesterol)
\item Adjusting the random intercept and slope mixed effects model for baseline CVD risk factors.
\end{enumerate}
Time-to-event models were adjusted for age, diabetes status, smoking status, baseline total cholesterol, baseline HDL cholesterol and sex.

For the Bayesian estimation we used diffuse uniform prior distributions $U[0,100]$ for standard deviations, uniform $U[-1,1]$ prior distributions for correlation parameters and diffuse normal prior distributions $N(0,100^2)$ for all other parameters. Priors were specified for the bivariate and trivariate normal distributions by expressing them as two and three conditional univariate normal distributions respectively. We used a burn-in of 1000 MCMC updates for the mixed effects models and 2000 MCMC updates for the joint models. We calculated results from 1000 sampled updates, and checked convergence using the Gelman-Rubin statistic as modified by Brooks and Gelman \cite{Brooks1998GeneralSimulations}. For the joint models there was a high degree of auto-correlation for the usual SBP and SBP variability hazard ratios, so MCMC chains were thinned by 4 updates.  

\afterpage{
\begin{figure}[t!]
	\centering
	\includegraphics[width=0.5\textwidth]{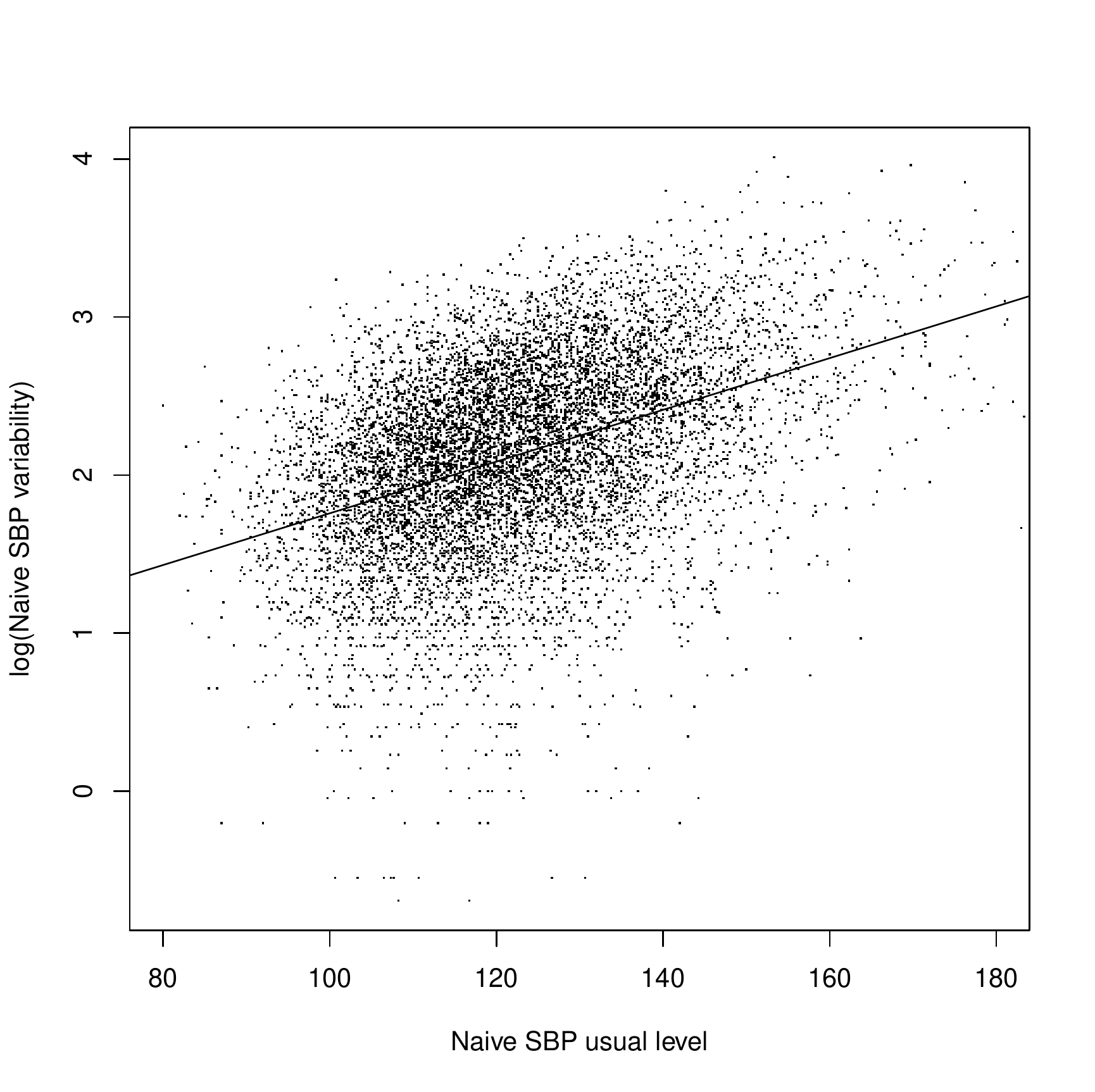} 
	\caption{Scatter plot of log naive SBP variability estimates against naive SBP usual level estimates with a fitted regression line.}
    \label{fig:cor}
\end{figure}
\newpage
}

Figure \ref{fig:hist} shows histograms of the estimated SBP usual level and SBP variability using the naive method. Also plotted are fitted normal and log-normal distributions respectively. The plots suggest that the assumption of a normal distribution for the random intercepts and a log-normal distribution for the residual SDs is appropriate. The naive usual level estimates are plotted against the naive variability estimates in Figure \ref{fig:cor}. There is a positive correlation with those with higher SBP tending to have greater SBP variability. The correlation between the naive estimates is 0.42, but the true correlation without measurement error is likely to be higher.

Estimates of usual levels and variabilities from the mixed effects models are compared with the naive estimates in Figure \ref{fig:naivevslmm}. The plotted lines are the lines of agreement between the two methods. Note that there is more difference between the methods for SBP variability estimates than for usual level estimates, with lower SBP variabilities being underestimated by the naive method and higher variabilities being overestimated.

\afterpage{
\begin{figure}[h!]
    \centering
        \subfloat[]{\includegraphics[width=0.47\textwidth]{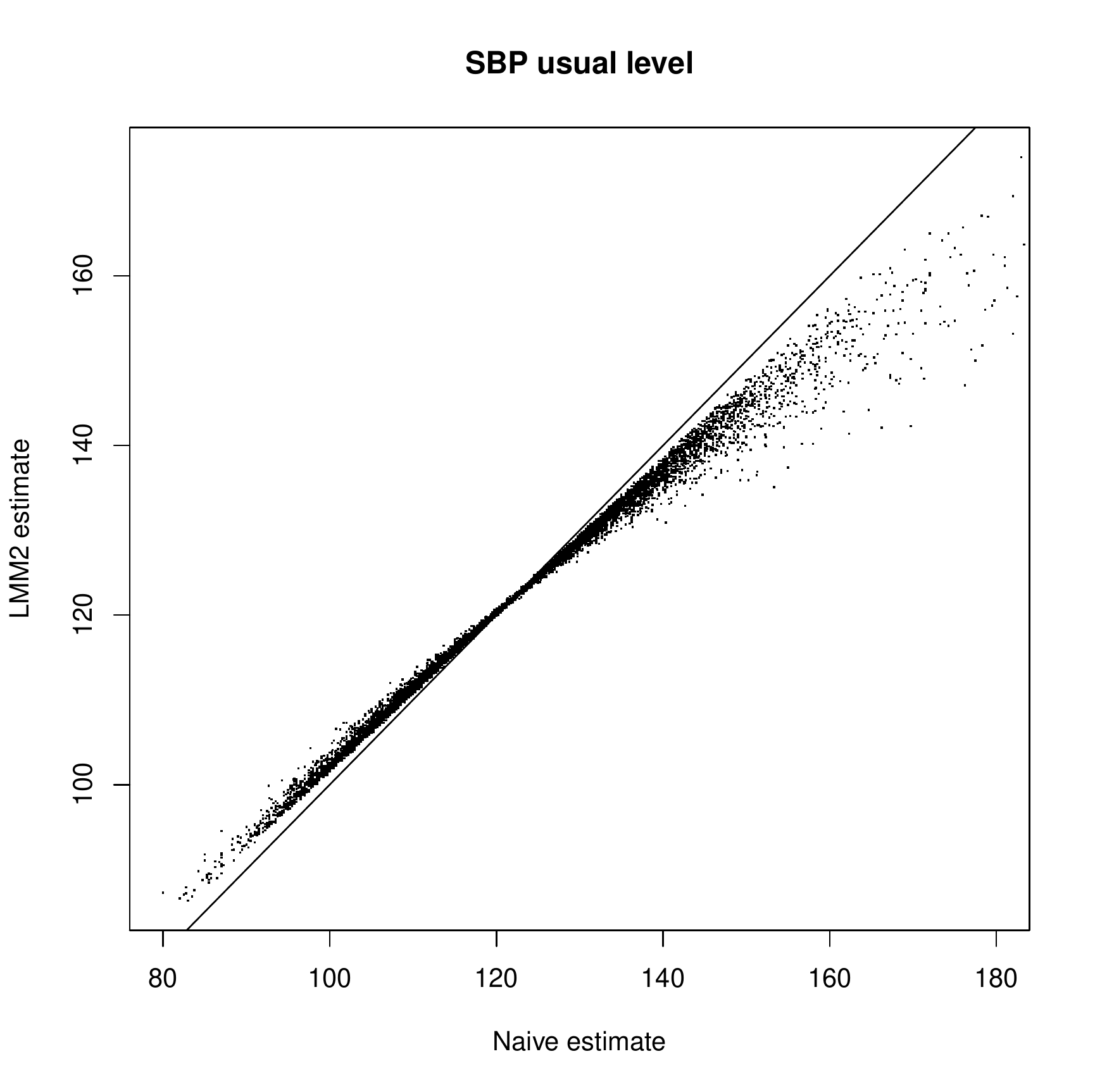}} \qquad
        \subfloat[]{\includegraphics[width=0.47\textwidth]{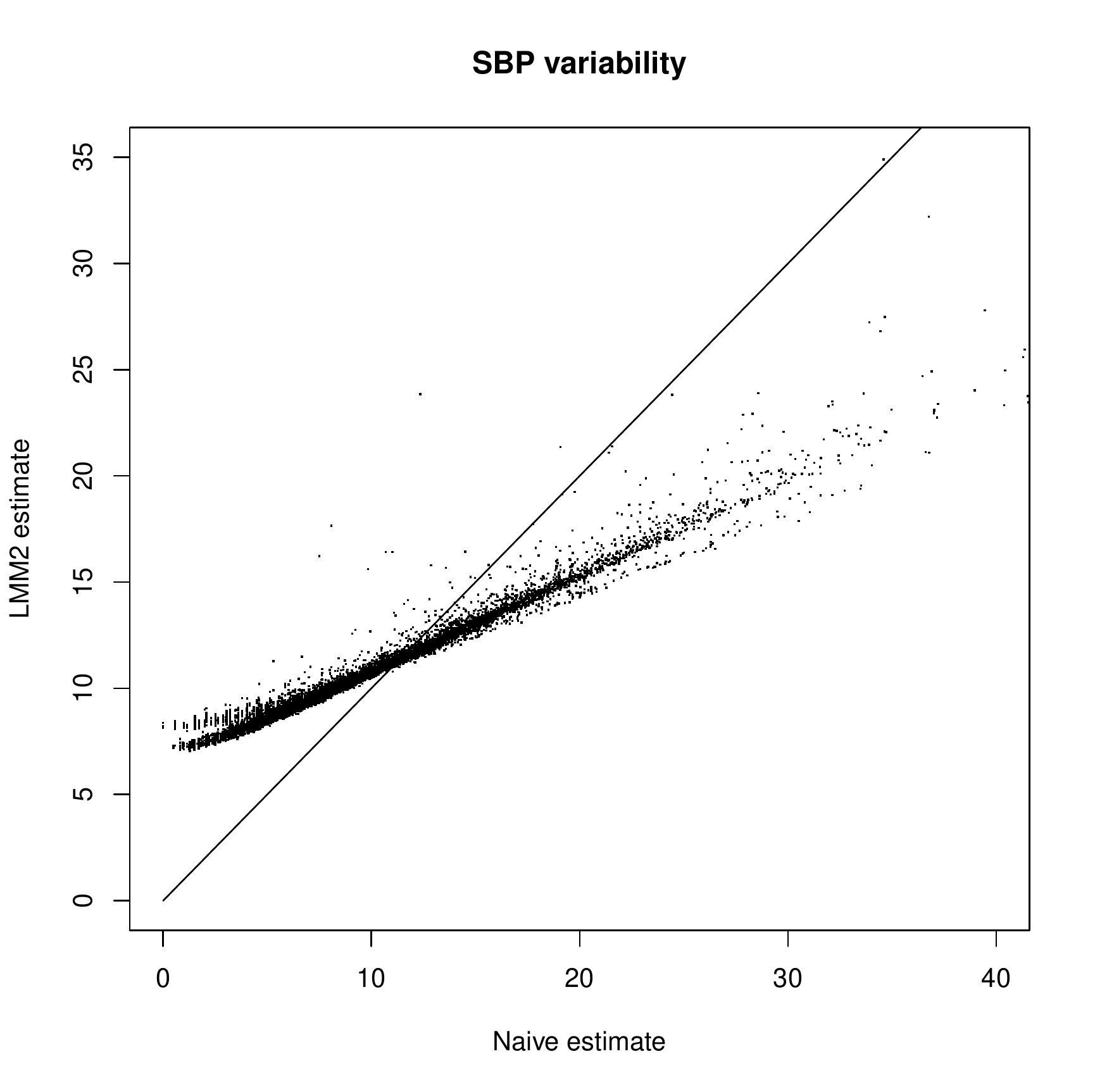}}
    \caption{Scatter plots of estimates using the LMM2 method against estimates using the naive method of (a) SBP usual level and (b) SBP variability with lines of agreement between the two estimates.}
    \label{fig:naivevslmm}
\end{figure}
}

Results from all models applied to the reduced ARIC data set are shown in table \ref{table:ARIC}. As expected, the joint models took considerably longer to run than the two-stage models with runtimes of between 17 and 20 hours compared to between 4 and 10 minutes respectively. All methods find an increased risk of CVD events with higher SBP variability. The methods which include random slopes find no evidence for an association between SBP gradient and CVD risk after adjusting for the usual level of SBP and SBP variability. 

The estimated association of SBP variability with CVD events is smaller for the naive model, suggesting regression dilution bias in this parameter estimate. Method LMM1, which does not account for the correlation between the SBP usual level and variability, gives a higher estimate of the association with the usual level of SBP than method LMM2, which does account for the correlation, as would be expected from the arguments given in Section 1 of the Supplementary Materials. The estimate of the association with SBP variability is generally lower for LMM2 than LMM1, suggesting a slight bias in the LMM1 estimates. The standard errors are, however, considerably larger for LMM2 than for LMM1. There is little difference between the results from the joint models and the results from the two-stage LMM models. Adjusting for other CVD risk factors in the mixed effects model makes little difference to the estimated association with SBP usual level and variability.   

Results from all models applied to the full ARIC data set are shown in table \ref{table:ARICfull}. For the models without random slopes the logHRs for the SBP usual level and SBP variability are similar to those from the truncated dataset. Standard errors are slightly smaller reflecting the increased sample size. But for the models with random slopes the logHRs for the effect of slope on the hazard of CVD are very different for the full dataset, with more positive slopes now leading to a lower risk of a CVD event when adjusted for the SBP usual level and SBP variability. This is counterintuitive to what we would expect; as higher SBP is associated with greater risk of CVD we would expect those with increasing SBP over time to also have a higher risk. 

\afterpage{
\newgeometry{top=0.7cm, bottom=0.7cm}
\begin{landscape}
\begin{table}[h]
\caption{Results (logHRs with standard errors in brackets) from all models applied to reduced data from the ARIC study. Methods LMM1, LMM2, LMM3 and LMM4 are two-stage methods respectively with (1) random intercept only and no correlation between the random intercept and the variability, (2) random intercept only and correlation, (3) random intercepts and slopes and no correlation between the random effects and the variability, (4)  random intercepts and slopes and correlation. JM1, JM2, JM3 and JM4 are the corresponding joint models. Adjusted LMM's adjust the longitudinal models/sub-models for baseline CVD risk factors. 
 \label{table:ARIC}}
\centering
\setlength{\tabcolsep}{4pt}
\begin{tabular}{llllllllll}
\toprule
\multicolumn{1}{l}{Model}&\multicolumn{1}{c}{Usual level $\alpha_1$}&\multicolumn{1}{c}{Variability $\alpha_\sigma$}&\multicolumn{1}{c}{Slope $\alpha_3$}&\multicolumn{1}{c}{Total chol.}&\multicolumn{1}{c}{HDL chol.}&\multicolumn{1}{c}{Age}&\multicolumn{1}{c}{Smoker}&\multicolumn{1}{c}{Diabetic}&\multicolumn{1}{c}{Male sex}\\ \midrule
\multicolumn{3}{l}{\bfseries Naive Model}&&&&&&&\\
~~&0.019 (0.002)&0.017 (0.005)&&0.114 (0.029)&-0.369 (0.085)&0.072 (0.006)&-0.503 (0.069)&-0.67 (0.084)&0.273 (0.067)\\
\multicolumn{3}{l}{\bfseries Unadjusted LMM}&&&&&&&\\
~~LMM1&0.023 (0.003)&0.045 (0.011)&&0.113 (0.029)&-0.37 (0.085)&0.072 (0.006)&-0.505 (0.069)&-0.671 (0.084)&0.273 (0.067)\\
~~LMM2&0.015 (0.006)&0.053 (0.021)&&0.113 (0.029)&-0.37 (0.085)&0.072 (0.006)&-0.506 (0.069)&-0.668 (0.084)&0.272 (0.067)\\
~~JM1&0.023  (0.003)&0.045 (0.011)&&0.114 (0.029)&-0.380 (0.087)&0.072 (0.006)&-0.498 (0.068)&-0.670 (0.086)&0.274   (0.065)\\
~~JM2&0.014 (0.007)&0.058 (0.023)&&0.114 (0.029)&-0.371 (0.085)&0.073 (0.006)&-0.503 (0.070)&-0.657 (0.081)&0.277 (0.068)\\
\multicolumn{3}{l}{\bfseries Adjusted LMM}&&&&&&&\\
~~LMM1&0.023 (0.003)&0.047 (0.011)&&0.129 (0.029)&-0.405 (0.085)&0.084 (0.006)&-0.447 (0.068)&-0.754 (0.083)&0.297 (0.067)\\
~~LMM2&0.012 (0.007)&0.062 (0.022)&&0.125 (0.029)&-0.402 (0.085)&0.081 (0.006)&-0.437 (0.068)&-0.749 (0.083)&0.313 (0.067)\\
~~JM1&0.023 (0.003)&0.045 (0.012)&&0.129 (0.028)&-0.408 (0.085)&0.085 (0.006)&-0.452 (0.066)&-0.766 (0.085)&0.297 (0.069)\\
~~JM2&0.010 (0.007)&0.071 (0.023)&&0.127 (0.028)&-0.404 (0.084)&0.081 (0.006)&-0.440 (0.070)&-0.754 (0.090)&0.318 (0.069)\\
\multicolumn{3}{l}{\bfseries Unadjusted LMM with slope}&&&&&&&\\
~~LMM3&0.023 (0.003)&0.047 (0.009)&0.024 (0.074)&0.112 (0.029)&-0.37 (0.085)&0.072 (0.006)&-0.507 (0.069)&-0.657 (0.084)&0.264 (0.067)\\
~~LMM4&0.016 (0.006)&0.053 (0.018)&0.01 (0.089)&0.113 (0.029)&-0.371 (0.085)&0.072 (0.006)&-0.507 (0.069)&-0.657 (0.084)&0.263 (0.067)\\
~~JM3&0.023 (0.003)&0.047 (0.011)&0.020 (0.077)&0.114 (0.027)&-0.370 (0.086)&0.073 (0.006)&-0.517 (0.072)&-0.659   0.088&0.265 (0.069)\\
~~JM4&0.015 (0.005)&0.056 (0.015)&0.004 (0.089)&0.114 (0.029)&-0.374 (0.087)&0.072 (0.006)&-0.512 (0.069)&-0.662   (0.085)&0.265 (0.068)\\ 
\multicolumn{3}{l}{\bfseries Adjusted LMM with slope}&&&&&&&\\
~~LMM3&0.022 (0.003)&0.049 (0.009)&0.033 (0.071)&0.129 (0.029)&-0.403 (0.085)&0.084 (0.006)&-0.442 (0.068)&-0.747 (0.084)&0.296 (0.067)\\
~~LMM4&0.014 (0.006)&0.057 (0.017)&0.006 (0.085)&0.124 (0.029)&-0.406 (0.085)&0.081 (0.006)&-0.437 (0.068)&-0.723 (0.084)&0.303 (0.067)\\
~~JM3&0.023 (0.003)&0.049 (0.010)&0.021 (0.071)&0.130 (0.029)&-0.408 (0.085)&0.084 (0.006)&-0.458 (0.067)&-0.766 (0.082)&0.292 (0.067)\\
~~JM4&0.014 (0.006)&0.060 (0.017)&-0.003 (0.101)&0.126 (0.029)&-0.406 (0.083)&0.081 (0.006)&-0.441 (0.067)&-0.744 (0.081)&0.304 (0.067)\\
\bottomrule  
\end{tabular}
\end{table}
\end{landscape}
\restoregeometry
}

\afterpage{
\newgeometry{top=0.7cm, bottom=0.7cm}
\begin{landscape}
\begin{table}[h]
\caption{Results (logHRs with standard errors in brackets) from all models applied to full data from the ARIC study. Methods LMM1, LMM2, LMM3 and LMM4 are two-stage methods respectively with (1) random intercept only and no correlation between the random intercept and the variability, (2) random intercept only and correlation, (3) random intercepts and slopes and no correlation between the random effects and the variability, (4)  random intercepts and slopes and correlation. JM1, JM2, JM3 and JM4 are the corresponding joint models. Adjusted LMM's adjust the longitudinal models/sub-models for baseline CVD risk factors.   \label{table:ARICfull}}
\centering
\setlength{\tabcolsep}{4pt}
\begin{tabular}{llllllllll}
\toprule
\multicolumn{1}{l}{Model}&\multicolumn{1}{c}{Usual level $\alpha_1$}&\multicolumn{1}{c}{Variability $\alpha_\sigma$}&\multicolumn{1}{c}{Slope $\alpha_3$}&\multicolumn{1}{c}{Total chol.}&\multicolumn{1}{c}{HDL chol.}&\multicolumn{1}{c}{Age}&\multicolumn{1}{c}{Smoker}&\multicolumn{1}{c}{Diabetic}&\multicolumn{1}{c}{Male sex}\\ \midrule
\multicolumn{3}{l}{\bfseries Naive Model}&&&&&&&\\
~~&0.02 (0.001)&0.016 (0.003)& &0.118 (0.02)&-0.461 (0.061)&0.06 (0.004)&-0.62 (0.046)&-0.8 (0.054)&0.395 (0.048)\\
\multicolumn{3}{l}{\bfseries Unadjusted LMM}&&&&&&&\\
~~LMM1&0.023 (0.002)&0.042 (0.007)& &0.124 (0.019)&-0.436 (0.057)&0.06 (0.004)&-0.634 (0.044)&-0.853 (0.051)&0.382 (0.045)\\
~~LMM2&0.015 (0.004)&0.053 (0.014)& &0.124 (0.019)&-0.437 (0.057)&0.06 (0.004)&-0.632 (0.044)&-0.849 (0.051)&0.381 (0.045)\\
~~JM1&0.024 (0.002)&0.045 (0.007)& &0.126 (0.019)&-0.440 (0.055)&0.060 (0.004)&-0.636 (0.044)&-0.855 (0.050)&0.391 (0.045)\\
~~JM2&0.012 (0.005)&0.065 (0.016)& &0.125 (0.020)&-0.445 (0.058)&0.061 (0.004)&-0.633 (0.045)&-0.862 (0.050)&0.392   (0.047)\\
\multicolumn{3}{l}{\bfseries Adjusted LMM}&&&&&&&\\
~~LMM1&0.023 (0.002)&0.043 (0.006)& &0.137 (0.019)&-0.453 (0.057)&0.073 (0.004)&-0.59 (0.044)&-0.956 (0.051)&0.409 (0.045)\\
~~LMM2&0.013 (0.004)&0.058 (0.013)& &0.134 (0.019)&-0.456 (0.057)&0.07 (0.004)&-0.579 (0.044)&-0.949 (0.051)&0.42 (0.045)\\
~~JM1&0.024 (0.002)&0.046 (0.007)& &0.139 (0.019)&-0.456 (0.061)&0.073 (0.004)&-0.603 (0.045)&-0.988 (0.051)&0.415 (0.045)\\
~~JM2&0.009 (0.005)&0.074 (0.016)& &0.137 (0.020)&-0.462 (0.058)&0.070 (0.004)&-0.575 (0.043)&-0.943 (0.050)&0.439 (0.049)\\
\multicolumn{3}{l}{\bfseries Unadjusted LMM with slope}&&&&&&&\\
~~LMM3&0.025 (0.002)&0.045 (0.006)&-0.129 (0.049)&0.123 (0.019)&-0.441 (0.057)&0.059 (0.004)&-0.635 (0.044)&-0.823 (0.051)&0.363 (0.045)\\
~~LMM4&0.02 (0.004)&0.048 (0.011)&-0.212 (0.06)&0.123 (0.019)&-0.442 (0.057)&0.059 (0.004)&-0.632 (0.044)&-0.821 (0.051)&0.359 (0.045)\\
~~JM3&0.026 (0.002)&0.049 (0.006)&-0.183 (0.066)&0.125 (0.019)&-0.453 (0.060)&0.060 (0.004)&-0.647 (0.046)&-0.845   (0.048)&0.366 (0.047)\\
~~JM4&0.018 (0.004)&0.061 (0.012)&-0.302 (0.085)&0.128 (0.019)&-0.458 (0.058)&0.061 (0.004)&-0.646 (0.047)&-0.847 (0.055)&0.374 (0.046)\\
\multicolumn{3}{l}{\bfseries Adjusted LMM with slope}&&&&&&&\\
~~LMM3&0.026 (0.002)&0.047 (0.006)&-0.118 (0.051)&0.138 (0.019)&-0.458 (0.057)&0.072 (0.004)&-0.599 (0.044)&-0.953 (0.051)&0.398 (0.045)\\
~~LMM4&0.019 (0.004)&0.051 (0.01)&-0.185 (0.06)&0.132 (0.019)&-0.461 (0.057)&0.069 (0.004)&-0.598 (0.044)&-0.96 (0.051)&0.39 (0.045)\\
~~JM3&0.027 (0.002)&0.052 (0.006)&-0.163 (0.061)&0.141 (0.019)&-0.469 (0.057)&0.074 (0.004)&-0.608 (0.046)&-0.968 (0.052)&0.405 (0.045)\\
~~JM4&0.017 (0.004)&0.065 (0.011)&-0.253 (0.083)&0.136 (0.019)&-0.480 (0.058)&0.070 (0.004)&-0.587 (0.045)&-0.939 (0.052)&0.416 (0.047) \\
\bottomrule  
\end{tabular}
\end{table}
\end{landscape}
\restoregeometry
}
\restoregeometry

\section{Discussion}
\label{sec:discussion}

We have proposed two-stage and joint modelling methods to estimate the association between visit-to-visit SBP variability and CVD, both of which use extensions of standard mixed effects models to allow different individuals to have different SBP variabilities. Both methods reduce regression dilution bias in the estimated hazard ratio compared to naive methods that have previously been used where SBP variability is estimated on an individual-by individual basis. In addition, the joint modelling methods allow for informative truncation of the repeated measurement process by the event of interest. In practice the method of choice will depend on the likely extent of the biases incurred by the two-stage approach vs the computational tractability of the joint modelling approach.

An important assumption of the two-stage method is that truncation of the observation process is non-informative, otherwise the number of observations may depend on the underlying risk level, which may lead to bias. Informative truncation can be avoided in the two-stage approach if distinct periods of follow-up for longitudinal and survival data can be defined, but this may only be possible at the cost of discarding information from individuals with events during longitudinal follow-up and repeated measurements taken during survival follow-up. In the clinical literature analyses have been conducted both using survival follow-up recorded subsequently to the repeated observations of SBP and using survival follow-up recorded concurrently with SBP measurements. As an example of the former, Rothwell et al explored the use of 7 compared with 10 SBP measurements, but commented that when 10 SBP measurements were used survival follow-up was shorter\cite{rothwell2010}.

Another important assumption of our method is that the measurement schedule is non-informative, which will not be satisfied in some studies. Some clinical studies investigating the association between SBP variability and CVD have used data which has not been collected for research purposes, such as data from electronic health records. For example, Gosmanova et al used data from around 3 million US veterans receiving healthcare from the US Veterans Health Administration\cite{GosmanovaAssociationDisease} and 
Hippisley-Cox et al used SBP variability as a risk predictor in the most recent QRISK tool for predicting CVD based on around 8 million individuals attending general practices in the UK\cite{Hippisley-Cox2017DevelopmentStudy}. In such cases additional bias may be incurred by informative observation of the repeated measurements, where the act of measurement is more (or less) likely in individuals at greater risk of disease. In these cases it may be necessary to extend the joint model to include an additional sub-model for the recurrent event process of an SBP measurement being taken. However, for cohort studies with a pre-specified measurement schedule, such as the ARIC study analyse in this paper, such additional model complexity is unlikely to be warranted. We therefore leave this investigation for future research.  

Results from our simulation studies suggest that the best model giving the least bias and coverage closest to nominal levels is the joint model which allows for correlation between the SBP usual level and the SBP variability. The naive method gives considerable bias in the association between SBP variability and CVD, as expected. Results previously reported in the CVD literature using the naive method could be subject to considerable bias in both the estimated association and its standard error. Our simulation results also indicate there may be a complex interplay between different sources of bias. While the use of mixed effects models and joint models can reduce regression dilution bias, model misspecification, such as not accounting for correlation between these variables, may introduce additional sources of bias.

Our analyses of the ARIC data found evidence of a positive association between higher SBP variability and the risk of CVD events. For this dataset results between the two-stage LMM approach and the joint modelling approach were similar, suggesting that the biases incurred by use of the two-stage approach were minimal in this example. However, use of the joint model comes at a cost of considerably longer computational time. This might be reduced by running multiple MCMC chains in parallel. If computational time is an issue then a possible modelling strategy would be to use the two-stage method for model selection and the joint model for inference. If there was also clinical interest in identifying explanatory variables associated with the longitudinal variability then the mixed effects models could be extended to incorporate a linear predictor in the mean of the $\log \sigma_i$ distribution, $\mu_{\sigma,i}=\zeta^T X^\sigma_i$. Similar models have previously been considered by Hedeker et al\cite{Hedeker2008AnData,Hedeker2012} and Goldstein et al\cite{GoldsteinMultilevelFunction}.

\section*{Acknowledgements}
We thank Angela Wood for helpful discussions. JKB was funded by the Medical Research Council grant numbers G0902100 and MR/K014811/1 and unit programme number MC\_UU\_00002/5. RP was funded by Medical Research Council grant number MR/N027485/1. This work was supported by the UK Medical Research Council (G0800270), British Heart Foundation (SP/09/002), UK National Institute for Health Research Cambridge Biomedical Research Centre, European Research Council (268834UK). The Atherosclerosis Risk in Communities study has been funded in whole or in part with
Federal funds from the National Heart, Lung, and Blood Institute, National Institutes of
Health, Department of Health and Human Services, under Contract nos.
(HHSN268201700001I, HHSN268201700003I, HHSN268201700005I,
HHSN268201700004I, HHSN2682017000021).
The authors thank the staff and participants of the ARIC study for their important
contributions.

\bibliographystyle{plain}
\bibliography{BP_variability}

\newpage
\setcounter{section}{0}
\begin{center}
\textbf{{\Large Supplementary Materials to `Estimating the association between blood pressure variability and cardiovascular disease: An application using the ARIC Study'}}
\end{center}

\begin{center}
\textbf{{ Jessica Barrett, Raphael Huille, Richard Parker, Yuichiro Yano and Michael Griswold}}\\
\vspace{0.1in}
\end{center}

\section{Bias caused by ignoring correlations in linear regression}
\label{sec:bias}
Consider a linear regression of outcome $y$ on correlated covariates $x_1$ and $x_2$,
\[
y = \beta_1 x_1 + \beta_2 x_2 + \varepsilon \,\, ,
\]
where $\varepsilon \sim N(0, \sigma_y^2)$ is the residual error in $y$. Now let
\[
x_1 = \rho \frac{\sigma_1}{\sigma_2} x_2 + \lambda \,\, ,
\]
where $\lambda$ and $x_2$ are uncorrelated, $\textrm{Var}(x_1)=\sigma_1^2$, $\textrm{Var}(x_2)=\sigma_2^2$ and $\rho$ is the correlation between $x_1$ and $x_2$. Re-writing the linear predictor,
\[
y = \beta_1 \lambda + \left( \rho \frac{\sigma_1}{\sigma_2} \beta_1 + \beta_2 \right) x_2 \,\, .
\]
As $\lambda$ and $x_2$ are uncorrelated we find
\begin{equation}
\hat{\beta}_2 = \frac{\textrm{Cov}(y, x_2)}{\sigma_2^2} - \rho \hat{\beta}_1\frac{\sigma_1}{\sigma_2}
\label{eq:beta2}
\end{equation}
If $x_1$ and $x_2$ are estimated ignoring the correlation between them, effectively setting $\rho=0$, we would therefore expect to observe a positive bias in $\hat{\beta}_2$ when $\rho$ and $\hat{\beta}_1$ are both positive, because the second term in equation (\ref{eq:beta2}) is ignored. By symmetry, we would also expect to observe a positive bias in  $\hat{\beta}_1$ for positive $\rho$ and positive $\hat{\beta_2}$.


\section{Additional simulation studies} 

In this section we present the results of further simulation studies investigating performance of the models for different levels of association between the longitudinal trajectories and the CVD event and for different levels of correlation between the usual level and the variability for Scenario 1.

Table \ref{table:alpha} shows results for different values of the true logHR's $\alpha_0$ and $\alpha_\sigma$. All results are given for datasets with $n=4$ measurements per individual. When $\alpha_\sigma=0$, i.e. there is no association between the longitudinal variability and the time-to-event, the naive method has negative bias in the usual level logHR due to regression dilution, but there is no bias in the variability logHR. By contrast, for methods LMM1 and JM1 there is slight positive bias in the variability logHR because the correlation between the usual level and the variability has been ignored. In this case, therefore, using the mixed effects model may lead to increased bias compared to the naive method because the model is misspecified. Results for larger values of the logHRs show substantial bias in both logHRs for the naive and LMM methods, with the naive method performing the worst. The joint models give little bias in the logHR for the usual level, but some negative bias for the variability logHR. Coverage probabilities are very much lower than the nominal 95\% level, with those of the joint models being substantially closer than all other methods. These results suggest that for substantially larger effect sizes, or equivalently for substantially less variation in the usual levels or variabilities in the population, all methods require a greater number of measurements per individual to sufficiently reduce estimation bias.

Results for various values of the correlation parameter $\rho$ are shown in Table \ref{table:rho}. Again, each individual has $n=4$ measurements. When there is negative correlation we find negative bias in both association parameters for the naive, LMM1 and JM1 methods, as we would expect from the arguments given in Section \ref{sec:bias}. Bias in the usual level logHR using the methods LMM1 and JM1, which ignore the correlation, increases with increasing $\rho$. But for the variability logHR the pattern is less clear because of the interplay between regression dilution bias and the bias incurred by ignoring correlation in the mixed effects model. For models LMM2 and JM2, which account for the correlation, we found minimal bias in estimated effects and coverage close to nominal values in all cases.

\newpage
\newgeometry{left=1cm, right=1cm}

\begin{table}[h]
\caption{Scenario 1 results with different  levels of association between the usual level and the variability of the longitudinal outcome and the time-to-event. Presented are the true values, mean (standard deviation) of estimated log hazard ratios, root mean squared error and coverage over 1000 simulated datasets. Methods of analysis are (1) True values, where the true usual levels and variabilities are used as covariates, (2) Naive method, (3) LMM1, the two-stage approach with no correlation between the usual level and the variability, (4) LMM2, the two-stage approach with correlation, (5) JM1, the joint model with no correlation and (6) JM2, the joint model with correlation. \label{table:alpha}}
\centering
\begin{tabular}{lllllcllll}
\hline
\multicolumn{1}{l}{\bfseries  }&\multicolumn{4}{c}{\bfseries Usual level logHR \boldmath$\alpha_0$}&\multicolumn{1}{c}{\bfseries }&\multicolumn{4}{c}{\bfseries Variability logHR \boldmath$\alpha_\sigma$}\\
\multicolumn{1}{l}{}&\multicolumn{1}{c}{True}&\multicolumn{1}{c}{Mean (SD)}&\multicolumn{1}{c}{RMSE}&\multicolumn{1}{c}{Coverage}&\multicolumn{1}{c}{}&\multicolumn{1}{c}{True}&\multicolumn{1}{c}{Mean (SD)}&\multicolumn{1}{c}{RMSE}&\multicolumn{1}{c}{Coverage}\\ \hline
\multicolumn{3}{l}{\bfseries \boldmath$\alpha_0=0.02$, $\alpha_\sigma=0$}&&&&&&&\\
~~True values&0.02&0.0201 (0.0043)&0.0043&0.949&&0&-0.0011 (0.0138)&0.0139&0.955\\
~~Naive&0.02&0.0175 (0.0038)&0.0045&0.893&&0&0.0012 (0.0103)&0.0103&0.946\\
~~LMM1&0.02&0.0193 (0.0042)&0.0043&0.943&&0&0.0062 (0.0165)&0.0176&0.935\\
~~LMM2&0.02&0.0199 (0.005)&0.005&0.944&&0&-1e-04 (0.0192)&0.0192&0.953\\
~~JM1&0.02&0.0194 (0.0042)&0.0043&0.942&&0&0.0043 (0.017)&0.0175&0.943\\
~~JM2&0.02&0.0201 (0.0051)&0.0051&0.944&&0&-0.0024 (0.0198)&0.0199&0.956\\
\multicolumn{3}{l}{\bfseries \boldmath$\alpha_0=0.02$, $\alpha_\sigma=0.05$}&&&&&&&\\
~~True values&0.02&0.02 (0.0043)&0.0043&0.94&&0.05&0.0501 (0.0117)&0.0117&0.94\\
~~Naive&0.02&0.0208 (0.0037)&0.0038&0.947&&0.05&0.0284 (0.0094)&0.0235&0.31\\
~~LMM1&0.02&0.0227 (0.0042)&0.005&0.892&&0.05&0.0531 (0.0151)&0.0154&0.929\\
~~LMM2&0.02&0.0198 (0.0049)&0.0049&0.948&&0.05&0.0511 (0.0175)&0.0175&0.931\\
~~JM1&0.02&0.023 (0.0042)&0.0052&0.883&&0.05&0.0521 (0.0155)&0.0156&0.939\\
~~JM2&0.02&0.02 (0.0051)&0.0051&0.949&&0.05&0.0502 (0.0183)&0.0183&0.936\\
\multicolumn{3}{l}{\bfseries \boldmath$\alpha_0=0.05$, $\alpha_\sigma=0.02$}&&&&&&&\\
~~True values&0.05&0.0502 (0.0043)&0.0043&0.948&&0.02&0.0201 (0.0112)&0.0112&0.942\\
~~Naive&0.05&0.044 (0.0037)&0.007&0.6&&0.02&0.0125 (0.009)&0.0117&0.834\\
~~LMM1&0.05&0.0491 (0.0041)&0.0042&0.94&&0.02&0.036 (0.0139)&0.0212&0.767\\
~~LMM2&0.05&0.0495 (0.0049)&0.0049&0.94&&0.02&0.0216 (0.0161)&0.0161&0.944\\
~~JM1&0.05&0.0501 (0.0043)&0.0043&0.946&&0.02&0.0339 (0.0145)&0.02&0.843\\
~~JM2&0.05&0.0505 (0.0051)&0.0051&0.945&&0.02&0.0193 (0.0168)&0.0168&0.952\\
\multicolumn{3}{l}{\bfseries \boldmath$\alpha_0=0.1$, $\alpha_\sigma=0.25$}&&&&&&&\\
~~True values&0.1&0.1002 (0.0043)&0.0043&0.94&&0.25&0.251 (0.0106)&0.0106&0.952\\
~~Naive&0.1&0.082 (0.004)&0.0184&0.002&&0.25&0.0972 (0.0097)&0.1531&0\\
~~LMM1&0.1&0.0911 (0.0043)&0.0099&0.386&&0.25&0.1879 (0.0152)&0.0639&0.008\\
~~LMM2&0.1&0.0818 (0.0048)&0.0188&0.027&&0.25&0.1759 (0.0183)&0.0763&0.007\\
~~JM1&0.1&0.1081 (0.0055)&0.0098&0.695&&0.25&0.2191 (0.0191)&0.0363&0.589\\
~~JM2&0.1&0.0969 (0.0057)&0.0065&0.905&&0.25&0.2096 (0.0215)&0.0458&0.503\\
\hline  
\end{tabular}
\end{table}

\begin{table}[h!]
\caption{Scenario 1 results with different levels of correlation between the usual level and the variability of the longitudinal outcome. Presented are the true values, mean (standard deviation) of estimated log hazard ratios, root mean squared error and coverage over 1000 simulated datasets. Methods of analysis are (1) True values, where the true usual levels and variabilities are used as covariates, (2) Naive method, (3) LMM1, the two-stage approach with no correlation between the usual level and the variability, (4) LMM2, the two-stage approach with correlation, (5) JM1, the joint model with no correlation and (6) JM2, the joint model with correlation.\label{table:rho}}
\centering
\begin{tabular}{lllllcllll}
\hline
\multicolumn{1}{l}{\bfseries  }&\multicolumn{4}{c}{\bfseries Usual level logHR \boldmath$\alpha_0$}&\multicolumn{1}{c}{\bfseries }&\multicolumn{4}{c}{\bfseries Variability logHR \boldmath$\alpha_\sigma$}\\
\multicolumn{1}{l}{}&\multicolumn{1}{c}{True}&\multicolumn{1}{c}{Mean (SD)}&\multicolumn{1}{c}{RMSE}&\multicolumn{1}{c}{Coverage}&\multicolumn{1}{c}{}&\multicolumn{1}{c}{True}&\multicolumn{1}{c}{Mean (SD)}&\multicolumn{1}{c}{RMSE}&\multicolumn{1}{c}{Coverage}\\ \hline
{\bfseries \boldmath$\rho=-0.5$}&&&&&&&&&\\
~~True values&0.02&0.02 (0.0043)&0.0043&0.939&&0.05&0.0497 (0.0127)&0.0127&0.957\\
~~Naive&0.02&0.0148 (0.0038)&0.0064&0.739&&0.05&0.025 (0.01)&0.0269&0.255\\
~~LMM1&0.02&0.0165 (0.0042)&0.0055&0.883&&0.05&0.0399 (0.0163)&0.0192&0.916\\
~~LMM2&0.02&0.0199 (0.0048)&0.0048&0.94&&0.05&0.0507 (0.019)&0.019&0.951\\
~~JM1&0.02&0.0167 (0.0043)&0.0054&0.894&&0.05&0.0382 (0.0164)&0.0202&0.898\\
~~JM2&0.02&0.0201 (0.005)&0.005&0.937&&0.05&0.0493 (0.0192)&0.0192&0.955\\
{\bfseries \boldmath$\rho=0$}&&&&&&&&&\\
~~True values&0.02&0.0198 (0.0038)&0.0038&0.946&&0.05&0.0496 (0.0105)&0.0105&0.951\\
~~Naive&0.02&0.0177 (0.0036)&0.0042&0.892&&0.05&0.0293 (0.0087)&0.0224&0.309\\
~~LMM1&0.02&0.0196 (0.0039)&0.004&0.945&&0.05&0.0509 (0.0147)&0.0148&0.947\\
~~LMM2&0.02&0.0196 (0.0039)&0.004&0.946&&0.05&0.0509 (0.0148)&0.0148&0.94\\
~~JM1&0.02&0.0199 (0.004)&0.004&0.943&&0.05&0.0496 (0.015)&0.015&0.955\\
~~JM2&0.02&0.0199 (0.004)&0.004&0.941&&0.05&0.0497 (0.015)&0.015&0.949\\
{\bfseries \boldmath$\rho=0.2$}&&&&&&&&&\\
~~True values&0.02&0.0201 (0.0037)&0.0037&0.95&&0.05&0.0501 (0.0105)&0.0105&0.959\\
~~Naive&0.02&0.0191 (0.0035)&0.0036&0.949&&0.05&0.0294 (0.0088)&0.0224&0.32\\
~~LMM1&0.02&0.021 (0.0039)&0.004&0.949&&0.05&0.0525 (0.0147)&0.0149&0.935\\
~~LMM2&0.02&0.02 (0.004)&0.004&0.954&&0.05&0.0509 (0.0151)&0.0151&0.939\\
~~JM1&0.02&0.0213 (0.0039)&0.0042&0.945&&0.05&0.0514 (0.0151)&0.0152&0.94\\
~~JM2&0.02&0.0202 (0.004)&0.004&0.956&&0.05&0.0497 (0.0154)&0.0154&0.949\\
{\bfseries \boldmath$\rho=0.5$}&&&&&&&&&\\
~~True values&0.02&0.02 (0.0043)&0.0043&0.94&&0.05&0.0501 (0.0117)&0.0117&0.94\\
~~Naive&0.02&0.0208 (0.0037)&0.0038&0.947&&0.05&0.0284 (0.0094)&0.0235&0.31\\
~~LMM1&0.02&0.0227 (0.0042)&0.005&0.892&&0.05&0.0531 (0.0151)&0.0154&0.929\\
~~LMM2&0.02&0.0198 (0.0049)&0.0049&0.948&&0.05&0.0511 (0.0175)&0.0175&0.931\\
~~JM1&0.02&0.023 (0.0042)&0.0052&0.883&&0.05&0.0521 (0.0155)&0.0156&0.939\\
~~JM2&0.02&0.02 (0.0051)&0.0051&0.949&&0.05&0.0502 (0.0183)&0.0183&0.936\\
{\bfseries \boldmath$\rho=0.8$}&&&&&&&&&\\
~~True values&0.02&0.0201 (0.0055)&0.0055&0.959&&0.05&0.0503 (0.015)&0.015&0.955\\
~~Naive&0.02&0.0237 (0.0041)&0.0055&0.831&&0.05&0.0242 (0.0098)&0.0276&0.217\\
~~LMM1&0.02&0.0256 (0.0046)&0.0073&0.767&&0.05&0.0495 (0.015)&0.015&0.948\\
~~LMM2&0.02&0.0202 (0.0077)&0.0077&0.951&&0.05&0.0506 (0.0245)&0.0245&0.946\\
~~JM1&0.02&0.0259 (0.0047)&0.0075&0.749&&0.05&0.0485 (0.0154)&0.0155&0.953\\
~~JM2&0.02&0.02 (0.008)&0.008&0.949&&0.05&0.051 (0.026)&0.026&0.946\\
\hline  
\end{tabular}
\end{table}

\clearpage
\newpage
\restoregeometry
\section{ARIC study baseline characteristics}

Table \ref{table:baseline} shows baseline characteristics of individuals in the reduced and full ARIC datasets at the study baseline. The age and cholesterol levels are similar between the two datasets, but the reduced dataset has a lower proportion of males, smokers and diabetics than the full dataset. 

\begin{table}[h]
\caption{Baseline characteristics of individuals in the reduced and full ARIC datasets; mean (SD) unless stated otherwise. \label{table:baseline}}
\centering
\begin{tabular}{ccc}
\hline
& Reduced dataset & Full dataset\\
& n = 10,019 & n = 13,161 \\
\hline
Age, years & 54.1 (5.7) & 54.4 (5.7) \\
Sex (male), n (\%) & 4286 (42.8) & 5924 (45.0) \\
Total cholesterol, mmol/L & 5.5 (1.0) & 5.5 (1.1) \\
HDL cholesterol, mmol/L & 1.4 (0.4) & 1.4 (0.4) \\
Smoker, n(\%) & 2372 (23.7) & 3671 (27.9) \\
Diabetic, n(\%) & 817 (8.2) & 1407 (10.7) \\
\hline  
\end{tabular}
\end{table}

\end{document}